\documentstyle[11pt]{article}

\begin{document}
\title{Do galaxy mergers form elliptical galaxies? A comparison of
kinematic and photometric properties}
\author{Phil James$^1$, Cheryl Bate$^1$, Martyn Wells$^2$,\\ Gillian
Wright$^2$, Ren\'e Doyon$^3$\\
$^1$Astrophysics Research Institute, Liverpool John Moores University,\\
Twelve Quays House, Egerton Wharf, Birkenhead L41 1LD.\\
$^2$UK Astronomy Technology Centre, Royal Observatory,\\ 
Blackford Hill, Edinburgh EH9 3HJ.\\
$^3$Department de Physique, Universit\'e de Montr\'eal and Observatoire du\\ 
Mont Megantic C.P. 6128, Succ. A., Montr\'eal (Qc), Canada H3C 3J7
\\
email: {\tt paj} {\tt @astro.livjm.ac.uk}}
\maketitle
\begin{abstract}
We present near-IR K-band imaging and spectroscopy of a sample of galaxy
mergers, which we use to derive light profile indices, absolute
magnitudes and central velocity dispersions.  It is found that the light
distributions of mergers more nearly resemble those of ellipticals than
of bulges, but that the mergers lie well away from the Fundamental Plane
defined by the ellipticals.  This is interpreted as being due to
enhancement of the K-band surface brightness of the mergers by a
significant population of supergiant stars, and independent evidence for
such a population is inferred from measurements of the depth of the
2.3~$\mu$m CO absorption feature.

\end{abstract}


\section{INTRODUCTION}
  
\noindent
The fate of mergers of gas-rich galaxies is an enduring question, which
has attracted significant observational and theoretical interest. Such
mergers must happen, and are probably common in at least some galaxy
environments; but opinion is sharply divided on the end-product of this
process. This is a possible mechanism for producing elliptical galaxies,
but it has also been argued that diffuse, dynamically cold disk-dominated
systems cannot become highly centrally-concentrated elliptical galaxies,
and that the end-product of a disk-disk merger is likely to be something
much more flimsy and diffuse (Ostriker 1980). Given this uncertainty, it
is clearly of interest to study the properties of systems that are
generally acknowledged as merger remnants, to determine their likely
fates.  This is the purpose of the present paper, with the ultimate aim
of this programme being to investigate a possible evolutionary sequence
between mergers and elliptical galaxies.

There are problems with using optical imaging and spectroscopy
to study the dynamics of mergers.  Mergers are extremely dusty systems,
with visual extinctions of $\sim$10 magnitudes towards their nuclei in
the early stages of the merging process, and
the dust tends to be in a very irregular distribution.  This causes great
uncertainties in the interpretation of optical measurements of these
systems, since the optical emission must come from an irregular outer
shell of the galaxy which represents the most disturbed and unrelaxed
component.  This is particularly unfortunate for studies of the
progression of the relaxation process in mergers, since relaxation will
first occur in the central region where the characteristic timescales are
shortest.  These regions can in principle be observed in HI 21 cm
emission, which is unaffected by the extinction problem, but many
galaxies are known to have central deficiencies of atomic gas, and in
very disturbed systems it is far from certain that the dynamics of the
gas component bear any relation to the stellar component we are
interested in here.

The near-infrared thus provides the best techniques for measuring
directly the kinematics of the stellar component in the presence of
extinction.  K-band (2.2$\mu$m) light is affected by extinction an order
of magnitude less than blue light, and thus can probe deep into the
centres of these galaxies.  This light is dominated by the old stellar
population, at least in normal galaxies.  Rix and Rieke (1993) find that
in the disks of star-forming galaxies, there is a significant
contribution from red supergiants, on the order of 15-25\% of the K-band
light, but this is very much less than the contribution of young stars to
the optical emission in such systems.  We consider the effects of
supergiant emission in the K band for an extreme starburst system later
in the paper. It is fortunate also that the K band contains the strong CO
absorption feature which is well suited to the measurement of velocity
dispersions of galaxies (Doyon et al. 1994b; Lester and Gaffney 1994;
Gaffney, Lester \& Doppmann 1995). Since this is a photospheric
absorption, it traces directly the kinematics of the stellar population.
Thus, by combining K band photometry with CO spectroscopy, it is possible
to study the structure and kinematics of the stellar population in the
central regions of merging galaxies.  This is the aim of the present
work.

\section{PREVIOUS WORK}

Many authors have contributed to the literature on this question.  Toomre
and Toomre (1972) made the cautious suggestion that dynamical friction
and consequent merging of pairs of disk galaxies ``seems like a scenario
for nothing less than the delayed formation of some elliptical galaxies-
or at least of major stellar halos from otherwise gas-rich disks.'' Many
numerical simulations of the merging process have been performed (e.g.
Toomre and Toomre 1972; White 1979; Farouki and Shapiro 1982; Barnes
1988), and have almost unanimously concluded that the R$^{1/4}$ profile
typical of luminous elliptical galaxies is the natural outcome of the
merging process.  This has received some observational support, with both
optical (Schweizer 1982) and near-infrared (Wright et al. 1990; Stanford
and Bushouse 1991) imaging showing the apparent emergence of such
profiles in ongoing mergers.  There are some problems with the
merger/elliptical scenario, however.  The phase-space argument of
Ostriker (1980) remains unresolved, although the discovery of strong
central concentrations of molecular gas in mergers (Sanders et al. 1986)
and subsequent nuclear starbursts (Joseph and Wright 1985) may contribute
the extra stellar density needed to resolve this discrepancy. The
apparent deficit of globular clusters in spirals relative to ellipticals
also remains problematic, with the suggestion that globulars may be made
during the merging process (Schweizer 1986) now being seriously
challenged (Whitmore and Schweizer 1995).  It is also possible that the
R$^{1/4}$ profiles in some or all of the mergers are due to pre-existing
bulge components, and may not be indicative of the light profile of the
entire system when relaxation has completed.

While this paper was in preparation, a very similar study was published by Shier and
Fischer (1998).  They also used the CO absorption feature to determine
central velocity dispersions, in their case for a sample of 10 galaxy
mergers.  In general, their conclusions are very similar to ours, and the
results will be compared throughout the present paper.

\section{GALAXY SAMPLE}

\noindent
Galaxies were chosen on the basis of a variety of visual morphological
criteria, which are generally accepted as being indicative of disk galaxy
mergers.  The most important of these is the presence of extended, narrow
tails which are found in N-body simulations to be a general feature of
galaxy mergers, and are caused by the strong tidal field generated during
the merging process.  The central regions of the mergers are visually
chaotic, due in large part to the disrupted dust lanes of the pre-merger
spirals.  Redshifts were restricted to $<$15000 kms$^{-1}$ to ensure that
the CO absorption edge, at a rest wavelength of 2.293$\mu$m, remained
well within the K atmospheric window.

\section{OBSERVATIONS}

\noindent
All observations presented here were made with the 3.8 metre United
Kingdom Infrared Telescope (UKIRT) on Mauna Kea, Hawaii.  K-band imaging
was obtained on the 3 nights 1990 April 30 - May 2, using the near- IR
camera IRCAM, which employed a 62$\times$58 pixel InSb array.  The 60~mm
focal length camera was used, giving a pixel scale of 1.24~arcsec.  All
the observations were made in the standard K filter.  Exposure times were
typically 10-15 seconds on chip before reading, to give
sky-background-dominated noise, co-added to give 30-40 minutes total
exposure per galaxy. Equal time was spent on nearby sky positions to
build up sky flats.  Telescope pointings were jittered around the mean
position to increase the area observed around galaxies, and to enable the
sky flats to be median filtered, thus minimising the effect of foreground
stars.  Seeing was typically $\sim$1.5~arcsec, but was difficult to
estimate exactly due to the large pixel scale used.

High resolution spectroscopy at 2.3$\mu$m was obtained using the
spectrometer CGS4 on UKIRT on the 3 nights 1994 June 6-8.  At that time,
CGS4 used a 62$\times$58 pixel InSb array.  We used the high resolution
echelle grating and the long focal length camera with a 2-pixel wide
slit, giving a FWHM spectral resolution equivalent to 16~km~s$^{-1}$ and a
spectral range of 0.01~$\mu$m.  Given the small spectral range, two
grating positions were used for each galaxy, giving a total spectral
coverage of 2.285-2.305~$\mu$m in the galaxy's restframe.  For each
galaxy, between 15 and 20 integrations each with exposure times of $\sim$180
seconds were taken.  The telescope was nodded so as to slide the target
galaxy up and down the slit between integrations to facilitate sky
subtraction.  Observations were also made of A stars to enable
atmospheric absorption features to be divided out, and BS~4737, a K1~III
star, was observed to provide a template of the CO absorption
profile.

\section{DATA REDUCTION}

\subsection{Imaging}

\noindent
Imaging data reduction was standard, using the KAPPA and FIGARO packages.
All frames were bias- and dark-subtracted, using dark frames with the
same on-chip exposures as the science frames and scaled to the same
number of coadds.  Flat-fielding was performed using median sky flats
taken in blank sky areas adjacent to the galaxy concerned, and with the
same exposure time as the galaxy.  Attempts to improve the
signal-to-noise of the flats by including frames taken over a longer
period of time yielded lower-quality final images, probably due to
changes in the effective flat field pattern as a function of changing sky
colour. Each galaxy was observed in at least 4 `jittered' pointings, and
these were combined into mosaics after flat-fielding and sky subtraction
of the individual sub-arrays.  Photometric calibration was taken from
images of the standard stars GL~406, HD~77281, HD~84800, HD~105601,
HD~106965, HD~136754, HD~161903 and HD~162208, which showed an overall
photometric scatter of $\pm$0.04 mag, when the zero points for all three
nights are combined.  The galaxy images were calibrated using the zero
point from the standard nearest in time and airmass, and differential
airmass corrections were very small, typically 0.01 or 0.02~mag.  The
derived galaxy light profiles for NGC~2623 and Arp~220 were in excellent
agreement with those in Wright et al. (1990), which had been obtained in
a previous run with IRCAM on UKIRT.  All K-band magnitudes and surface
brightnesses were corrected for redshifting of the passband and for
surface brightness dimming according to the prescription of Glazebrook et
al. (1995) who derive a single correction formula applicable to all
Hubble types.  This correction has a maximum value of --0.11~mag. for the
present sample, and a mean of --0.06~mag. For the present paper, the
final stage of data reduction was to obtain multi-aperture photometry
with circular software apertures, regardless of the ellipticity of the
galaxies, for comparison with the similar measures taken by Mobasher et
al. (1999) to derive log(D$_K$) values (described in the section 6.3
below) for their sample of ellipticals. We did this for the mergers by
using the Starlink routine APERADD, centring the apertures using the
CENTROID package with a sampling window of 9$\times$9 pixels.  Aperture
sizes were iterated to find the value within which the mean galaxy
surface brightness was 16.5 mag/square arcsec at K, as used by Mobasher
et al. (1999).  Column 4 of table 1 gives the derived log(D$_K$) values,
where D$_K$ is measured in arcsec, corrected to the distance of the Coma
Cluster for direct comparability with the Coma ellipticals.  This
correction was done assuming an unperturbed Hubble Flow and a Coma
Cluster redshift of 6942~km~s$^{-1}$~Mpc$^{-1}$. A Hubble constant of
75~km~s$^{-1}$~Mpc$^{-1}$ was assumed throughout.  Column 6 gives the log
of the effective diameter A$_e$, in arcsec and again corrected to the
Coma distance. A$_e$ (the diameter encompassing half the total light for
a pure R$^{1/4}$ profile) was approximated by a Petrosian diameter
(Petrosian 1976) with index $\eta =$ 1.39 (Kjaergaard 1993). $\eta$
is the difference between the surface brightness at a given radius and the
mean surface brightness within that radius.  This
definition of A$_e$ was used as it is insensitive to departures from pure
R$^{1/4}$ profiles, and should be relatively unaffected by fading in the
surface brightness of galaxies.  Column 7 gives the mean K-band surface
brightness within the diameter A$_e$.

The other data listed
in Table 1 are catalogued names (column 1), number in the Arp (1966) list
of peculiar and disturbed galaxies (column 2), total K-band absolute
magnitude (column 3), and best-fitting power-law index to the light
profile, as described in section 6.1 below (column 5).

\subsection{Spectroscopy}

\noindent
Initial data reduction was done at the telescope, using an automatic
reduction process to dark- and bias- subtract all frames, to flat-field
using internally-generated lamp flats, and to ratio by an A-star spectrum
to remove telluric absorptions.  The galaxies were slid up and down the
slit to facilitate sky-subtraction, and the processed
`Reduced Group' frames contained two spectra, one positive and one
negative, following the automatic subtraction of spectra taken at the two
positions.  Known bad pixels in the array were automatically set to magic
values, so that they would be ignored in subsequent reduction, but it was
also necessary to identify additional bad pixels and cosmic ray hits in
the Reduced Groups.  Wavelength calibration was done using an argon arc
lamp in CGS4.  No spectrophotometric calibration was attempted, as it was
not required for this project.

The positive and negative spectra were then extracted from each Reduced
Group frame and subtracted to give the final galaxy spectrum.  For the
template star BS~4737, a long baseline spectrum was obtained by observing
at 12 grating positions.  These spectra were combined using the ADJOIN
routine in FIGARO, giving the final stellar template spectrum.

The final stage of data reduction is to determine the velocity
dispersions in the galaxy mergers from the Doppler-smoothing of the CO
absorption. This absorption has an intrinsically steep edge at
2.293~$\mu$m which makes it sensitive to such effects. Following Doyon et
al. (1994b) and Gaffney et al. (1995) we assume that the velocity
smoothing function can be approximated by a Gaussian of the form $f(x) =
exp(-(x^2/2\sigma^2))$, where $x$ is displacement in velocity from the
systemic velocity of the galaxy and $\sigma$ is proportional to the width
of the Gaussian and hence the velocity dispersion.  To determine the
velocity dispersion, we convolved the template spectrum of BS~4737 with
Gaussians of different widths, and determined the best fit of the
resulting spectrum to the measured galaxy spectrum.  In each of these
fits, the free parameters were a multiplicative scaling of the spectral
fluxes, an additive constant to the spectral fluxes, and a wavelength
shift due to redshift differences between the galaxy and template star.
In each case, well-defined minima were found in the reduced chi-squared
of the fits with respect to each of the four free parameters.  Having
identified the best fit, errors were calculated for the width of the
velocity Gaussian by freezing the other parameters and changing the
Gaussian width until the reduced $\chi^{2}$ increased by the factors
corresponding to 1- and 2-$\sigma$ uncertainties.

Table 2 lists parameters derived from spectroscopic measurements. Column
1 gives the galaxy names, column 2 the catalogued redshifts, and column 3
the redshifts derived from the spectroscopic fits described above, quite
independently of the catalogued redshifts.  Good agreement is found with
the catalogued redshifts, giving confidence in the fitting process.
Column 4 of Table 2 gives the derived velocity dispersions with
1-$\sigma$ errors, and column 5 the spectroscopic CO index, which is
discussed further in section 6.3 below.

\section{ANALYSIS}

\subsection{Light profiles}

\noindent
We first address the question of whether the elliptical-like light light
profiles found for some mergers could be due to pre-existing bulges which
have survived the merging process. Figure 1 shows a comparison of the
best-fit indices of the K-band light profiles of the mergers (plotted as
circles) with the bulges of spirals (crosses) of Hubble types Sa--Sd
(Seigar and James 1998), and two ellipticals (squares), NGC~5845 and
NGC~6023. The fitted profiles are generalised de Vaucouleurs profiles,
and the index plotted is $(1/n)$ in the $R^{(1/n)}$ relation, such that a
true de Vaucouleurs profile would correspond to 0.25. The errors were
calculated by omitting 0, 1, 2 and 3 central points from the profiles and
redoing the fits. This excluded a circular area of radius up to
3.7~arcsec, significantly larger than the seeing size, and generally had
little impact on the measured indices. The spiral galaxy indices
generally lie between 0.4 and 1.0, with some evidence for a bimodal
distribution with $\sim$0.5 and $\sim$1.0 as preferred values. Similar
results were found by de Jong (1996).  However, the mergers show
significantly brighter total luminosities and cuspier light profiles,
with indices lying close to the value of 0.25, as do the two ellipticals
for which we have K-band imaging.  However, not all of the mergers
plotted are actually well-fitted by a de Vaucouleurs profile, with some
having significant ripples and irregularities in their profiles.  The
important result from Figure 1 is that even these systems show at least
the degree of central light concentration expected of an elliptical, and
their light distributions and luminosities are not those expected of
pre-existing spiral bulges, or of galaxy disks.  Indeed, there is even
some evidence that the mergers have lower indices and cuspier profiles
even than ellipticals, which may be evidence of an additional,
centrally-concentrated component in the K-band light, resulting from a
nuclear starburst.  This possibility is discussed further in section 6.3.

The referee commented that there appears to be a trend in profile index
with absolute magnitude, and suggests that the lower values found for the
mergers may simply reflect a selection bias, since the mergers are
significantly more luminous than the bulges, on average.  However, if
there is such an effect, it does not appear to be strong enough to cause
the result we find.  Regressing index on absolute magnitude, there is a
weak trend both for the bulges and mergers, in the sense that more
luminous systems have somewhat lower indices.  However, the correlations
have significances of less than 5\% in both cases. Extrapolating the
trend found for bulges to the mean absolute magnitude of the mergers
predicts that the average profile index for the mergers should be 0.51,
whereas the average of the measured indices is 0.23. Thus the mergers
genuinely seem to be offset to lower profile indices than bulges, even if
we were to compare with bulges of similar luminosity (which do not seem
to exist in any case).

The merger with the large error bar in Figure 1 is NGC~6052, which has a
highly irregular structure, even in K-band light.  Since the errors are
estimated from fitting the profile over different radial ranges, it would
appear that the unusual and highly irregular profile of NGC~6052 is
particularly sensitive to such changes.  It is clear that this galaxy
bears no resemblance to an elliptical galaxy, and appears quite different
from the other mergers.

\subsection{Velocity dispersions}

\noindent
Figure 2 shows the range of velocity dispersions found for ellipticals
and bulges by Whitmore, McElroy \& Tonry (1985).  These are split into
ellipticals (dashed line), lenticulars (solid line) and spirals (dotted
line).  The measured velocity dispersions for the mergers are also marked
at the top of Figure 2.  It can be seen that the majority of mergers have
velocity dispersions between 100 and 200 kms$^{-1}$, with a mean of
164$\pm$35 kms$^{-1}$, where the standard error on the distribution is
quoted.  This compares with mean velocity dispersions of 222$\pm$5
kms$^{-1}$ for ellipticals, 190$\pm$4kms$^{-1}$ for lenticular bulges,
and 142$\pm$4 kms$^{-1}$ for spiral bulges, from values in the Whitmore
et al. (1985) compilation. NGC~6090 has the lowest velocity dispersion of
any of the mergers, and this may well be related to the fact that it has
two widely-separated nuclei.  Thus for this galaxy we are almost
certainly seeing the velocity dispersion of one of the pre-existing
nuclei, and this value is likely to be significantly lower than the
disperson of the final, relaxed merger remnant.  At the other extreme,
NGC~6240 has a velocity dispersion larger than any of the spiral galaxy
bulges in the Whitmore et al. (1985) compilation, and the value is indeed
in the high-velocity tail of the distribution even for elliptical
galaxies. Again, this may not represent the final virialised velocity
dispersion of this system.  For example, there may be superposed
components with different systemic velocities along the line of sight,
which would be impossible to distinguish using the present data. Such
effects cannot be gross, since then the spectroscopy would resolve two
separate CO absorption features.  However, given the width of the CO
feature and limited signal-to-noise of the present data, no strong
constraints can be placed.

The distribution of velocity dispersions of the merger sample is shown by
a Kolmogorov-Smirnov test to be formally consistent with those of both
spiral bulges and ellipticals, and so our velocity data alone do not
permit any significant test of the dynamical evolution of mergers.  In
this, we differ somewhat from Shier and Fischer (1998), who conclude that
their velocity dispersion measurements are not consistent with the values
expected for L$_{\star}$ elliptical galaxies, since apart from NGC~6240,
the highest velocity dispersion they measure is 151~km~s$^{-1}$.

\subsection{Parameter correlations}

A strong test of the merger hypothesis is made possible by the parameter
correlations resulting from the `Fundamental Plane' of elliptical
galaxies (Dressler et al. 1987, Djorgovski \& Davis 1987), since we know
that any forming elliptical will have to arrive on this plane.  This
describes the tight 2-dimensional locus occupied by elliptical galaxies
in the 3-dimensional space defined by velocity dispersion, luminosity and
scale-size.  These latter two parameters can be conveniently grouped
together, and Dressler et al. (1987) found that a particularly useful
quantity was D$_n$, the diameter of a photometric aperture within which
the mean surface brightness of the galaxy has some value $n$.  Dressler
et al. took $n$ to be 20.75 mag. per square arcsec in the optical B
passband, and found a very tight correlation between the resulting D$_n$
values and velocity dispersion.  Mobasher et al. (1999) have examined the
equivalent relation (which they term the D$_K$--$\sigma $ relation) using
K-band photometry of elliptical galaxies in the Coma cluster, and find a
small scatter of 0.076 dex in the relation, similar to the optical
D$_n$--$\sigma$ relation. They also present a plot which explicitly shows
an edge-on projection of the K-band Fundamental Plane, thus reducing the
scatter even below that of the D$_K$--$\sigma$ relation. Here the plotted
quantities are 1.38log$\sigma+$0.3$<$SB$_K$$>_e+c_1$ vs logA$_e$, where
A$_e$ is the effective diameter and $<$SB$_K$$>_e$ the mean surface
brightness within this diameter. The dataset of Mobasher et al. thus
provides an excellent way to test whether our mergers obey the same
parameter correlations as do elliptical galaxies, since all the
parameters required for these relations are provided by our K-band
photometry and spectroscopy.

D$_K$ values were calculated using circular aperture photometry on the
K-band images, and finding the size in arcseconds of a circular aperture
which enclosed a mean surface brightness of 16.5~mag. per square arcsec.
This surface brightness, and the use of circular apertures, were
identical to the procedure followed by Mobasher et al. (1999), and so the
results should be directly comparable. Figure 3 shows the D$_K$--$\sigma$
relations for Coma ellipticals (crosses) and for the mergers (circles).
In both axes, the quantities actually plotted are logarithms to base 10.
The ellipticals show the expected correlation, whereas the mergers
scatter well above this locus.  This implies either that the mergers have
anomalously low velocity dispersions, or anomalously large D$_K$
diameters, or a combination of both, when compared with the Coma
ellipticals.

A similar result is shown in Figure 4, which compares the near-IR
Fundamental Plane of Coma ellipticals with the present merger sample.
Again the elliptical galaxies show a very tight correlation, whereas the
mergers scatter away from the Fundamental Plane in the same sense as was
found from Figure 3.  We will now discuss whether the offset is likely to
be due to anomalous velocity dispersions or surface brightnesses in the
mergers.

As noted above, the distribution of velocity dispersions of the merger
sample is consistent with that of elliptical galaxies (formally the means
of the two distributions differ by 1.6~$\sigma$ but the
Kolmogorov-Smirnov test shows the distributions to be indistinguishable),
and is unlikely to evolve further in these relaxed central regions. This
consistency is confirmed by Figure 3, with the possible exception of
NGC~6090, which may evolve to a higher velocity dispersion as the two
nuclei relax together. For the remainder of the mergers, any plausible
evolution onto the elliptical locus in Figure 3 seems likely to be
predominantly in the D$_K$ direction, and in every case would have to be
in the sense of making the galaxies become smaller, at the same mean
surface brightness level.  This is quite plausible, given that mergers
are widely acknowledged to be linked to highly luminous bursts of massive
star formation (e.g. Joseph \& Wright 1985; Barnes \& Hernquist 1991;
Solomon, Downes \& Radford 1992; but see also the counter arguments of
Thronson et al. 1990).  Such a starburst will result in a temporarily
enhanced luminosity and surface brightness of the system, and in the
K-band this is likely to result principally from emission from supergiant
stars. Thus we would argue that the most likely future evolution of the
mergers will shift them vertically in Figure 3, in the direction of
smaller D$_K$ values.  Assuming the starburst population to be well mixed
with the older stars, this fading will not affect the A$_e$ values, and
will shift points vertically upwards in Figure 4, again consistent with
moving them onto the Fundamental Plane defined by Coma ellipticals. We
will now estimate the likely size of this effect.

The photometric contribution of supergiants in mergers can be quantified
from the depth of the 2.3$\mu$m CO absorption feature, which is
characteristic of such stars (Doyon, Joseph \& Wright 1994a; Ridgway,
Wynn-Williams \& Becklin 1994; Goldader et al. 1995).  Using a method
described by Doyon et al. (1994a) it is possible to use low-resolution
K-band spectroscopy to make an estimate of the fraction of K-band light
contributed by the starburst population in a galaxy from the depth of the
CO feature.  This method assumes that the K-band light has contributions
from an old stellar population with a small CO index, and a starburst
population with a much larger CO index.  From the observed index, which
lies between these extremes, one can then solve for the fractional mix of
the two populations.  This simple form of population synthesis was
applied to the 5 mergers for which CO indices were available (Mkn~273,
Arp~220, NGC~6090 and NGC~6240 from Ridgway et al. 1994; IC~883 from
Goldader et al. 1995), assuming a spectroscopic CO index (see Doyon et
al. 1994a for a definition) of 0.2 for the old population and 0.34 for a
pure starburst.

It was found that the starburst fraction varied from providing only
$\sim$20\% of the K-band luminosity in Mkn~273, to $\sim$85\% in Arp~220.
We then made the simplifying assumptions that all of this starburst
population would ultimately disappear, and that this would uniformly
depress the surface brightness of the galaxy in the K-band.  It was then
straightforward to predict the effect of such fading on the D$_K$ values
for these 5 mergers, and the results are shown as arrows in Figure 3.
Clearly, this correction moves the points towards the elliptical locus,
and the movement is of the right order to explain the present
discrepancies.  The equivalent fading vectors in Figure 4 would shift the
points vertically upwards by between 0.1 and 0.6 in the quantity plotted
on the y axis, again of the correct order to shift points onto the
Fundamental Plane.  The corresponding arrows are not shown in this figure
to avoid confusing the plot.

A further test of this fading hypothesis is shown in Figure 5, where
log(A$_e$) is plotted against log($\sigma$).  Given the definition of
A$_e$ as a Petrosian diameter, it should be relatively unaffected by
fading, and certainly far less so than D$_K$ and $<$SB$_K$$>_e$.  Given
this, it is reassuring to see that the distribution of mergers is much
more consistent with that of ellipticals in this figure than in figures 3
and 4, and that the one very discrepant point, the double nucleus system
NGC~6090, is offset to lower log($\sigma$), as suggested earlier.

If there are red supergiants in the numbers necessary to explain the
observed CO indices, then for any plausible star formation history and
IMF it is to be expected that younger OB stars should also be present.
This is confirmed for the mergers plotted with arrows in Figure 3;
Goldader et al. (1997) find strong Brackett~$\gamma$ emission from all
five, consistent with the presence of OB stars.

\section{DISCUSSION}

A study of the dynamics of mergers was undertaken by  Lake \& Dressler
(1986), who compared the luminosity--velocity dispersion (L--$\sigma$)
relation with that of ellipticals (Faber \& Jackson 1976).  They combined
B and V photometry from the literature with Reticon spectroscopy of the
Ca triplet, and found a remarkably good
agreement between the distribution of velocity dispersions and optical
luminosities for mergers and ellipticals.  This result was argued to be
at variance with a simple dynamical model, which predicted that merging
should not lead to an increase in velocity dispersion.  Thus they
concluded that the merging process must involve significant dissipation
for the merger products to follow the L--$\sigma$ relation as found.

The L--$\sigma$ relation is very closely related to the D$_n$--$\sigma$
relation we have investigated. The apparent difference in results is
probably due to the merger sample selection. The only merger in common
between the two studies is NGC~7252, and this is possibly the most
evolved merger in our sample, but the youngest merger observed by Lake \&
Dressler.  The majority of their objects are recognisable as elliptical
galaxies, but with peculiarities such as strong shells and ripples, which
make it likely that these are mergers in the late stage of the relaxation
process, which is only beginning for most of our galaxies.  Thus it is
not surprising that different results are found by the two studies.

Shier and Fischer (1998) perform a K-band Fundamental Plane analysis of
their merger sample, like that of the present paper, and reach very
similar conclusions.  They also find that most of the mergers are
displaced from the plane defined by ellipticals, in the sense of being
bright for a given velocity dispersion.  From a population synthesis
argument, they estimate that the mergers are likely to fade by
1.5--2.0~mag in the K band during the next 3~Gyr, and that this fading
will place them on or near the elliptical locus, assuming no other scale
size or dynamical changes.

\section{CONCLUSIONS}

\noindent
The principal conclusions of this study are as follows.  We confirm that
in terms of their near-IR photometric properties, ongoing galaxy mergers
more nearly resemble elliptical galaxies than they do spiral bulges.
Even those poorly fitted by a de Vaucouleurs profile show a `cuspy' light
distribution which is likely to smooth into an elliptical-like profile
through relaxation processes.  Using velocity dispersions alone we are
unable to distinguish whether mergers more closely resemble bulges or
ellipticals, but this is unsurprising given the small number of mergers
in our sample and the extensive overlap between bulge and elliptical
$\sigma$ distributions.  A clear discrepancy is found between the
Fundamental Plane distributions and
D$_K$--$\sigma$ relations of mergers and ellipticals, however.  We
suggest that this may be due to a short-lived population of supergiant
stars temporarily increasing the surface brightness of the mergers, and
find evidence for such a population at approximately the predicted level
from literature measurements of the CO absorption strength.

It is clear that significant improvements on
these results are attainable with existing and planned instrumentation.
An intriguing possibility is that it may be possible, by combining
samples of young and old mergers, to map out an evolutionary sequence
from the highly distorted and dusty systems to the relaxed,
old-star-dominated elliptical galaxies, from their positions in the
Fundamental Plane parameter space.  Key issues here would be the
relative importance of star formation,  stellar evolution effects,
dynamical relaxation and dissipation in the formation of the Fundamental
Plane.

\section{ACKNOWLEDGMENTS}

\noindent
PJ thanks Sue Wild for many useful comments on a draft of this paper.
The referee is thanked for a careful reading of the paper and several
helpful suggestions which improved the content and clarity of presentation.
The United Kingdom Infrared Telescope is operated by the Joint Astronomy
Centre on behalf of the U.K. Particle Physics and Astronomy Research
Council. This research has made use of the NASA/IPAC Extragalactic
Database (NED) which is operated by the Jet Propulsion Laboratory,
California Institute of Technology, under contract with the National
Aeronautics and Space Administration.

\section{REFERENCES}

\noindent
Arp H., 1966, ApJS, 14, 1  

\noindent
Barnes J.E., 1988, ApJ, 331, 699

\noindent
Barnes J.E., Hernquist L.E., 1991, ApJ, 370, L65

\noindent
de Jong R. S., 1996, A\&A, 313, 45

\noindent
Djorgovski S., Davis M., 1987, ApJ, 313, 59

\noindent
Doyon R., Joseph R.D., Wright G.S., 1994a, ApJ, 421, 101

\noindent
Doyon R., Wells M., Wright G.S., Joseph R.D., Nadeau D., James P.A.,
1994b, ApJ, 437, L23

\hangindent=6mm \hangafter=1
\noindent
Dressler A., Lynden-Bell D., Burstein D., Davies R.L., Faber S.M.,
Terlevich R., Wegner G., 1987, ApJ, 313, 42

\noindent
Faber S.M., Jackson R.E., 1976, ApJ, 204, 668

\noindent
Farouki R.T., Shapiro S.L., 1982, ApJ, 259, 103

\noindent
Gaffney N.I., Lester D.F., Doppmann G., 1995, PASP, 107, 68

\noindent
Glazebrook K., Peacock J.A., Miller L.A., Collins C.A., 1995, MNRAS, 275,
169

\noindent
Goldader J.D., Joseph R.D., Doyon R., Sanders D.B., 1995, ApJ, 444, 97

\noindent
Goldader J.D., Joseph R.D., Doyon R., Sanders D.B., 1997, ApJS, 108, 449

\noindent
Joseph R.D., Wright G.S., 1985, MNRAS, 214, 87

\noindent
Kjaergaard P., Jorgensen I., Moles M., 1993, ApJ, 418, 617

\noindent
Lake G., Dressler A., 1986, ApJ, 310, 605

\noindent
Lester D.F., Gaffney N.I., 1994, ApJ, 431, L13

\noindent
Mobasher B., Guzm\'an R., Arag\'on-Salamanca A., Zepf S., 1999, 
MNRAS, 304, 225
 
\noindent
Ostriker J.P., 1980, ComAp, 8, 177

\noindent
Petrosian V., 1976, ApJ, 209, L1

\noindent
Ridgway S.E., Wynn-Williams C.G., Becklin E.E., 1994, ApJ, 428, 609

\noindent
Rix H.-W., Rieke M.J., 1993, ApJ, 418, 123

\hangindent=6mm \hangafter=1
\noindent
Sanders D.B., Scoville N.Z., Young J.S., Soifer B.T., Schloerb F.P., Rice
W.L., Danielson G.E, 1986, ApJ, 305, 45

\noindent
Schweizer F., 1982, ApJ, 252, 455

\noindent
Schweizer F., 1986, in Faber, S.M., ed., Nearly Normal Galaxies.
Springer-Verlag, Berlin, p.18

\noindent
Seigar M.S., James P.A., 1998, MNRAS, 299, 672

\noindent
Shier L. M., Fischer J., 1998, ApJ, 497, 163

\noindent
Solomon P.M., Downes D., Radford S.J.E., 1992, 387, L55

\noindent
Stanford S.A., Bushouse H.A., 1991, ApJ, 371, 92

\noindent
Thronson H.A., Majewski S., Descartes L., Hereld M., 1990, ApJ, 364, 456

\noindent
Toomre A., Toomre J., 1972, ApJ, 179, 623

\noindent
White S.D.M., 1979, MNRAS, 189, 831

\noindent
Whitmore B.C., McElroy D.B., Tonry J.L., 1985, ApJS, 59, 1

\noindent
Whitmore B.C., Schweizer F., 1995, AJ, 109, 960

\noindent
Wright G.S., James P.A., Joseph R.D., McLean I.S., 1990, Nature, 344, 417

\newpage

\begin{table*}
\caption{Galaxy names and photometric parameters}
\begin{center}
\begin{tabular}{lcccccc}
\hline
Galaxy Name & Arp No. & M$_K$ & Log(D$_K$) & Profile index & Log(A$_e$) &
$<$SB$_K$$>_e$\\

\hline
IC 883   & 193 & -24.05 & 1.226 & 0.36$\pm$0.05   & 1.21$\pm$0.05 & 16.44\\
IC 4553  & 220 & -24.46 & 1.322 & 0.24$\pm$0.09   & 1.22$\pm$0.03 & 16.10\\
Mkn 231  & --  & -27.08 & 1.903 & 0.17$\pm$0.05   & 1.00$\pm$0.05 & 14.77\\
Mkn 273  & --  & -25.19 & 1.497 & 0.14$\pm$0.05   & 1.24$\pm$0.02 & 15.48\\
NGC 2623 & 243 & -23.85 & 1.207 & 0.10$\pm$0.05   & 0.94$\pm$0.09 & 15.47\\
NGC 3509 & 335 & -24.24 & 1.008 & 0.025$\pm$0.05  & $>$1.59 & $>$18.3\\
NGC 4194 & 160 & -23.04 & 1.093 & 0.20$\pm$0.05   & 0.65$\pm$0.03 & 14.71\\
NGC 6052 & 209 & -23.44 & 0.923 & 0.83$\pm$0.5    & 1.36$\pm$0.01 & 17.58\\
NGC 6090 & --  & -24.49 & 1.331 & (0.05$\pm$0.05) & 1.35$\pm$0.03 & 16.53\\
NGC 6240 & --  & -25.85 & 1.640 & 0.11$\pm$0.05   & 1.55$\pm$0.03 & 16.17\\
NGC 7252 & 226 & -24.57 & 1.372 & 0.28$\pm$0.05   & 1.21$\pm$0.02 & 15.90\\
\hline 
\end{tabular}
\normalsize
\end{center}
\end{table*}

\begin{table*}
\caption{Spectroscopic parameters}
\begin{center}
\begin{tabular}{lcccc}
\hline
Galaxy Name& Catalogued recession  & Derived Recession	  & Velocity & CO$_{sp}$\\
	   & velocity (kms$^{-1}$) & Velocity (this work)& Dispersion & \\
\hline
IC 883   & 7000       & 7040  & 206$\pm$90     & 0.25$^2$ \\
IC 4553  & 5434       & --    & 150$\pm$21$^3$ & 0.32$^1$ \\
Mkn 231  & 12651      & --    & --             & -0.02$^1$ \\
Mkn 273  & 11326      & 11308 & 160$\pm$60     & 0.23$^1$ \\
NGC 2623 & 5535       & --    & --             & 0.24$^1$ \\
NGC 3509 & 7704       &--     & --             & -- \\
NGC 4194 & 2506       & 2523  & 104$\pm$25     & -- \\
NGC 6052 & 4716       & --    & --             & -- \\
NGC 6090 & 8795--9062 & 8929  & 50$\pm$20      & 0.30$^1$ \\
NGC 6240 & 7339       & --    & 359$\pm$21$^3$ & 0.29$^1$ \\
NGC 7252 & 4688       & 4743  & 123$\pm$19     & -- \\
\hline 
\end{tabular}
\end{center}
\hspace*{1.0cm}$^1$ Ridgway et al. 1994\\
\hspace*{1.0cm}$^2$ Goldader et al. 1995\\
\hspace*{1.0cm}$^3$ Doyon et al. 1994b\\
\end{table*}

\clearpage

\noindent
{\bf Figure 1.} Best-fitting profile index vs K-band absolute magnitude
for spiral bulges (crosses), ellipticals (squares) and  mergers (circles).

\vspace*{0.4cm}
\noindent
{\bf Figure 2.} Distribution of velocity dispersions taken from Whitmore
et al. (1985), showing ellipticals (dashed line), lenticular bulges
(solid line), and spiral bulges (dotted line).  The measured velocity
dispersions for the mergers are indicated at the top of the figure.

\vspace*{0.4cm}
\noindent
{\bf Figure 3.} The D$_K$--$\sigma$ relation for Coma ellipticals
(crosses), taken from Mobasher et al. (1999) and for the present sample
of mergers (circles).  The arrows show the effect of K-band fading
inferred from CO indices as explained in section 6.3.

\vspace*{0.4cm}
\noindent
{\bf Figure 4.} The Fundamental Plane relation for Coma ellipticals
(crosses), taken from Mobasher et al. (1999) and for the present sample
of mergers (circles). 

\vspace*{0.4cm}
\noindent
{\bf Figure 5.} Log(A$_e$) vs Log($\sigma$) for Coma ellipticals
(crosses), taken from Mobasher et al. (1999) and for the present sample
of mergers (circles). 

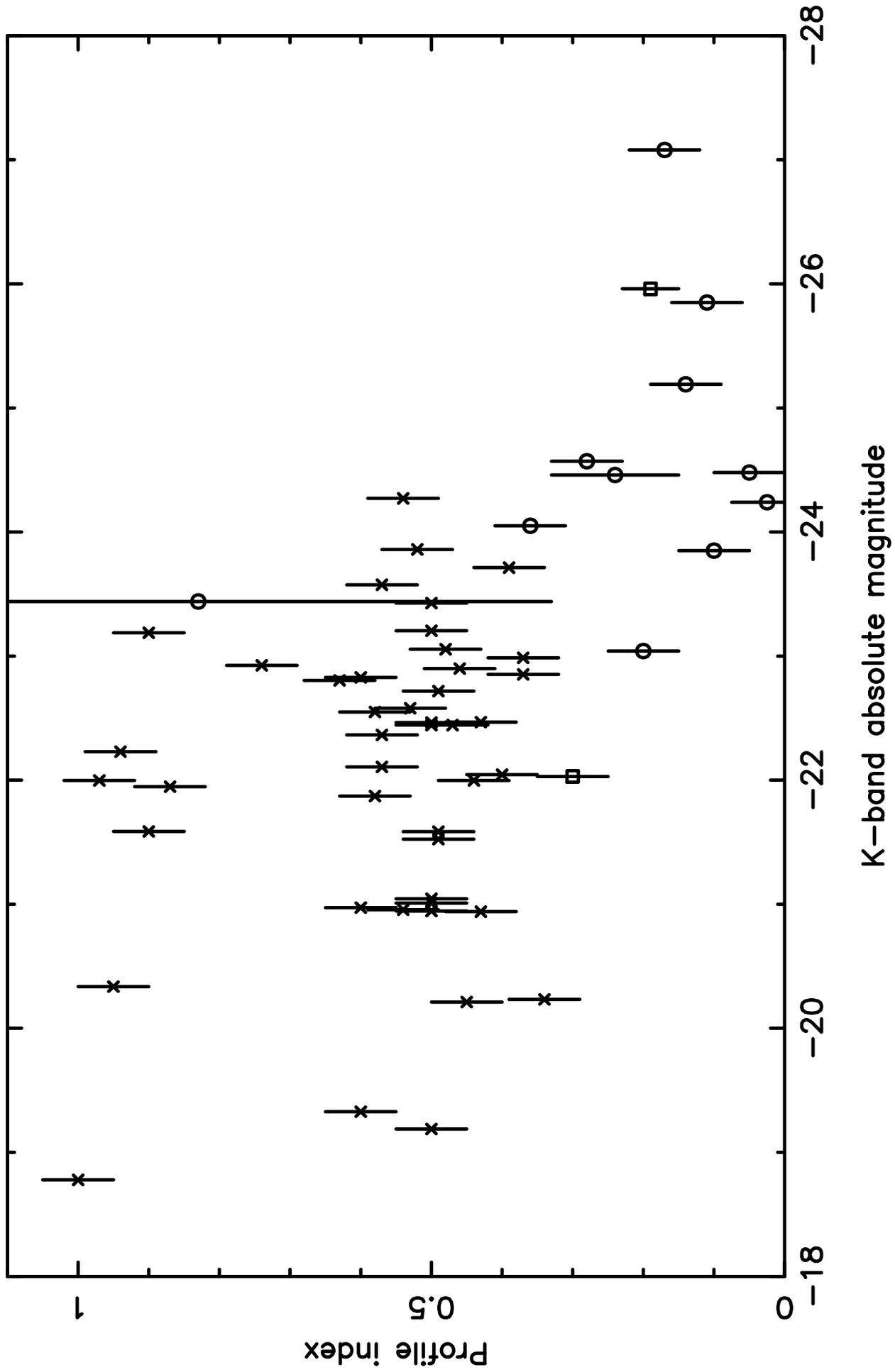
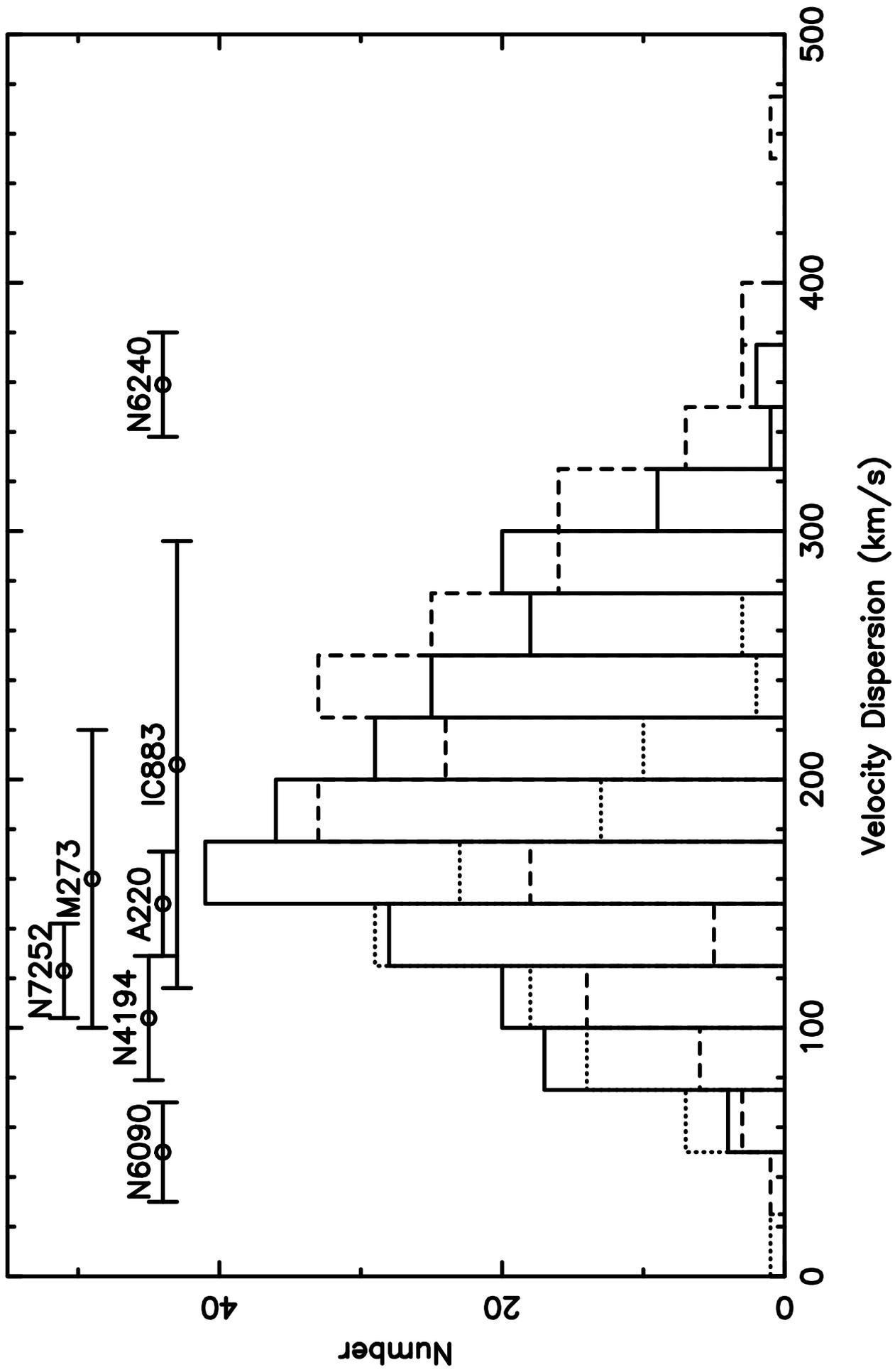
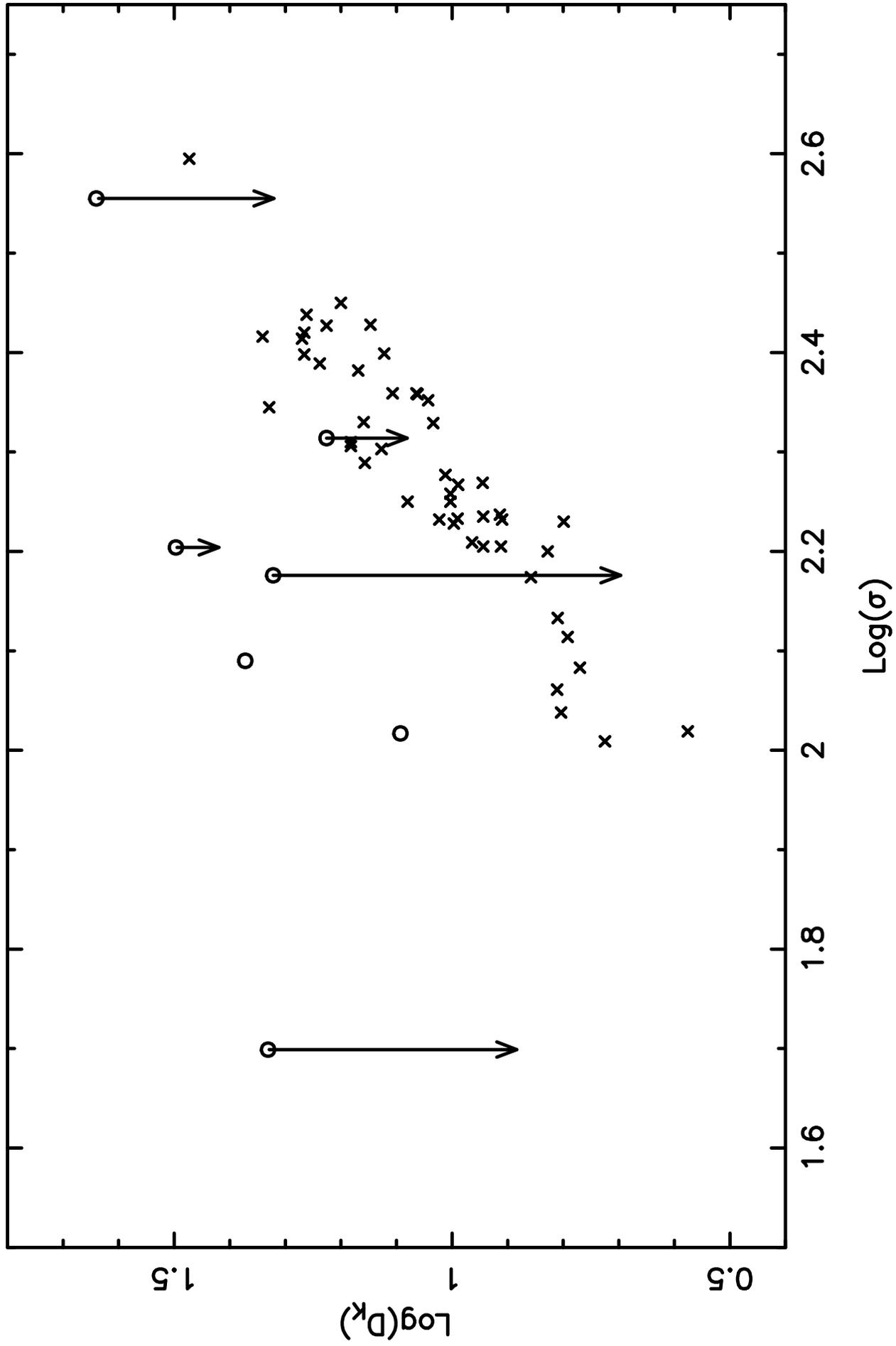
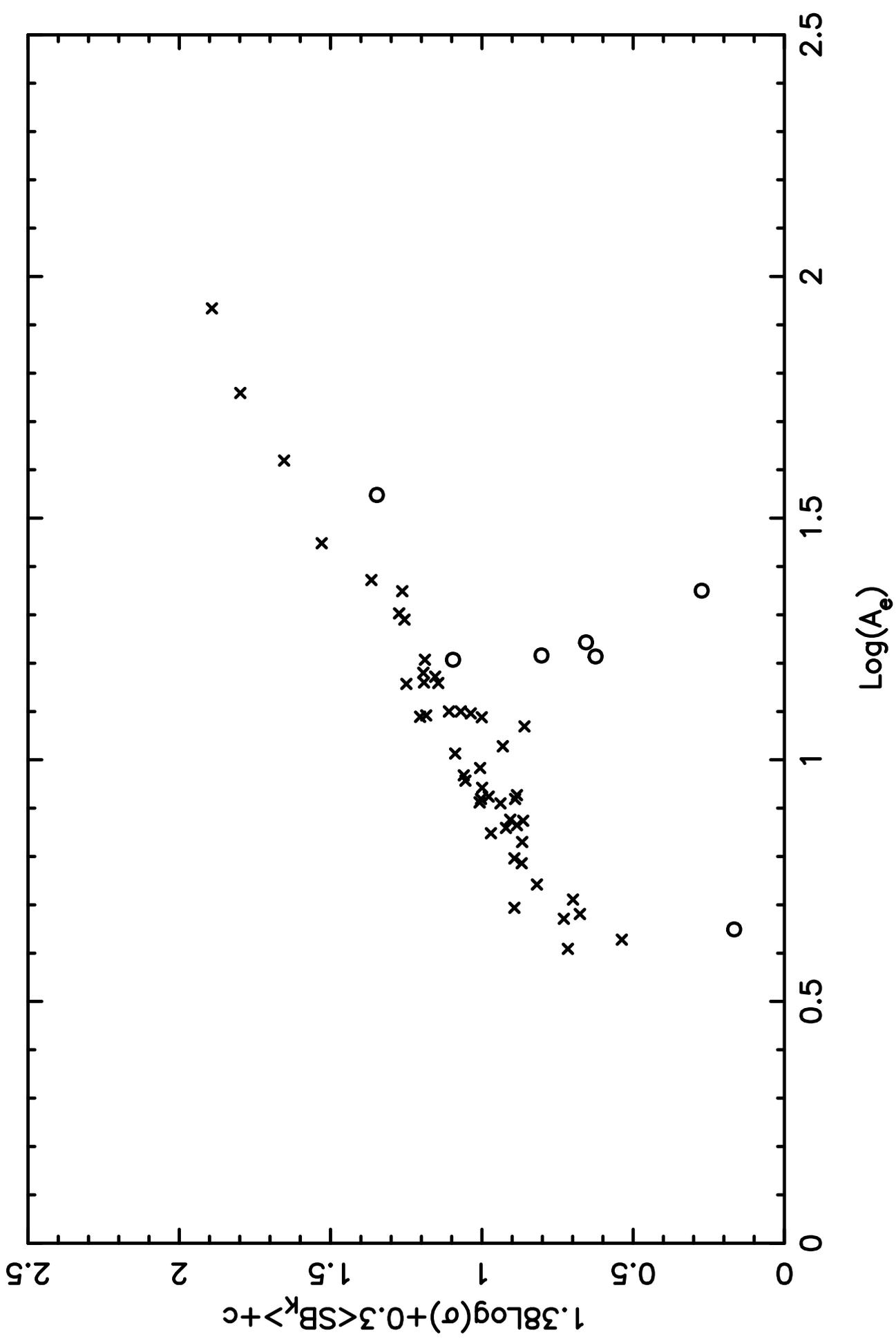
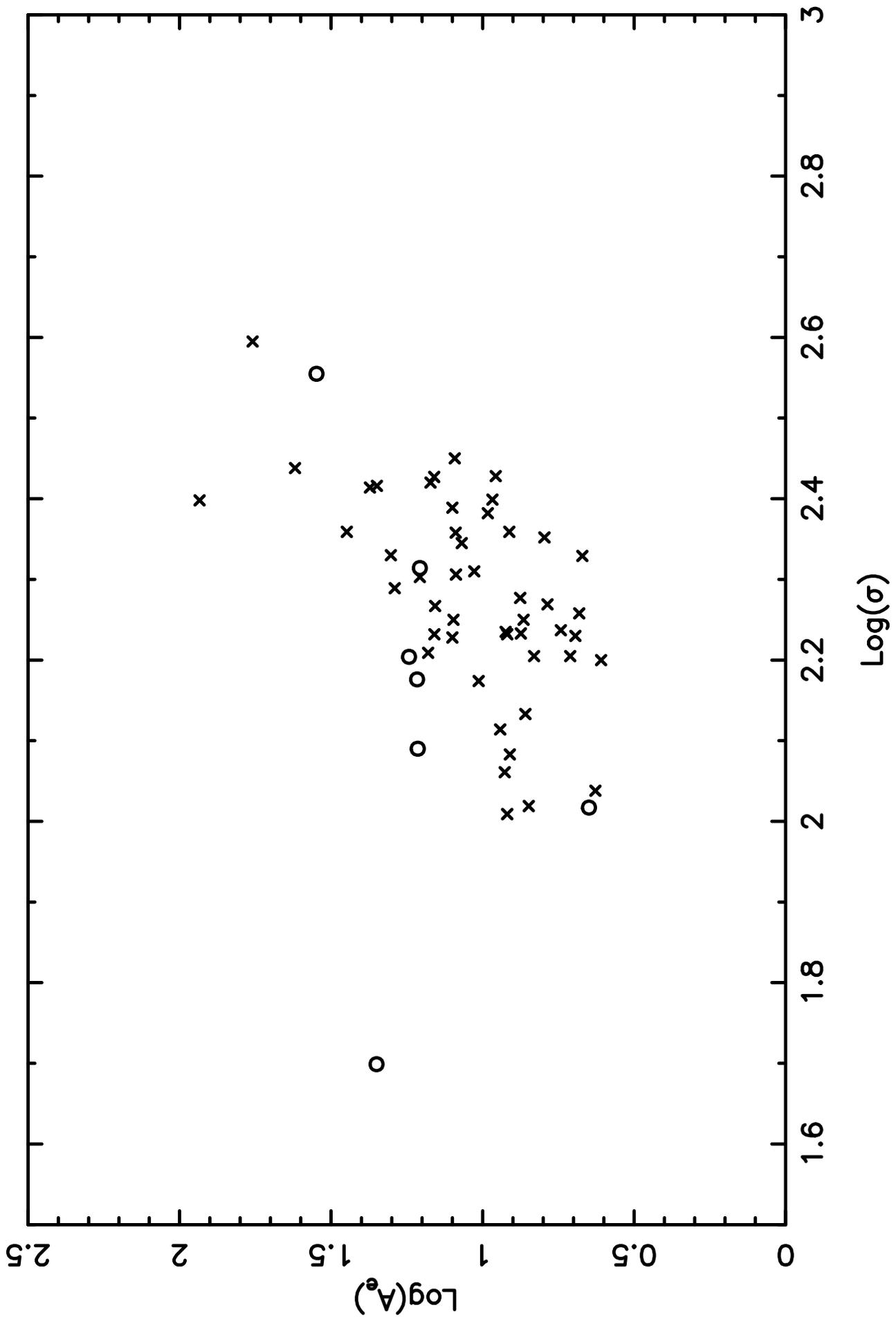
\end{document}